\documentclass[12pt,pra,aps,amssymb,amsmath,tightenlines,showpacs]{revtex4}
\usepackage{graphicx}

\newcommand{\half}{\mbox{$\textstyle \frac{1}{2}$}}

\newcommand{\rd}{\mbox{$\rm d$}}
\newcommand{\bea}{\begin{eqnarray}}
\newcommand{\eea}{\end{eqnarray}}
\newcommand{\cura}{\left(\begin{array}{c} 1 \\ 1 \end{array} \right)}
\newcommand{\curb}{\left(\begin{array}{c} 1 \\ -1 \end{array} \right)}

\begin{document}

\title{State reduction dynamics in a simplified QED model}

\author{Daniel~J.~Bedingham}
\email{d.bedingham@imperial.ac.uk}
\affiliation{Blackett~Laboratory,~Imperial College,~London~SW7~2BZ,~UK.}

\date{\today}

\begin{abstract}
A simplified model of quantum electrodynamics involving a charged two-state system 
interacting with an electromagnetic field mode is examined. By extending the 
Schr\"odinger equation to include stochastic and nonlinear terms the dynamical process of 
quantum state reduction can be represented. A specific choice of modified Schr\"odinger 
dynamics is shown to result in stable coherent field states. The two-state system 
undergoes an induced state reduction to a generalised current state due to its 
interaction with the field mode. Numerical results are presented demonstrating 
state reduction dynamics for an initial superposition of two current states. An induced 
reduction time-scale for the two-state system is derived and confirmed by the numerics.
\end{abstract}


\maketitle

\section{Introduction}

Standard quantum dynamics cannot describe how a particular measured outcome is realised in an 
experiment---this feature underlies our failure to unify classical physics with its quantum roots. 
A direct response to this problem is to take the view that standard quantum theory is incomplete and that 
a complete quantum theory should provide a dynamical description of quantum state reduction. In this way 
it is hoped to provide an objective description of quantum measurement without recourse to a classical 
measuring device.
   
The great success of quantum theory does not leave much room to manoeuvre when proposing
alternatives. Any new theory must agree with the experimentally verified predictions of quantum theory. 
This roughly implies that the dynamics should reduce to Schr\"odinger 
dynamics for micro systems and that state reduction effects should become more important with 
system size. More specifically, the form of any modifications we make should involve a random factor (at some level of
approximation) to reflect the random nature of state reduction, and we would expect some nonlinearity 
in the dynamics, reflecting a feedback from the state vector to the probability of the outcome \cite{pearlorig0}. 

These features can be adequately represented by a stochastic differential equation for the 
quantum state \cite{pearlorig,gisin} (for general reviews see \cite{Bass, Pear2}), the most well known model 
of this type being the continuous spontaneous localisation (CSL) model \cite{ghir2}. In the CSL model, state 
reduction occurs onto a basis of smeared position states leading to definite localised positions for systems of 
macroscopic size. However, there is scope for further model building, principally because we can choose from many
other preferred bases for state reduction (e.g.~\cite{adle,Dorj,bass2}). Until we have observational evidence of state reduction
dynamics we are free to experiment with different proposals.

Here we investigate the state reduction process for a simplified form of quantum electrodynamics.
We consider a charged two-state quantum system interacting with a quantum electromagnetic field mode.
We investigate a model which describes state reduction onto a coherent field state basis. 
State reduction for the two-state system happens indirectly via standard interaction with the field 
mode. We find that the two-state system evolves towards a generalised current eigenstate.

The appealing feature of this model is that the stable eigenstates of the system are a natural description 
of the classical world whereby the field is represented by a coherent state and the two-state system
(representative of charged matter) by a current eigenstate. Modifications to Schr\"odinger's equation
only involve the field mode---it is not necessary to make further modifications in order to independently 
describe state reduction in the two-state system. Also, it is possible to generalise the model to field theory 
without generating infinite particle creation \cite{me} (this is a problem which appears, e.g.,~with the 
CSL model \cite{pear3,ghir1}), although it should be pointed out that this results in a loss of Lorentz 
invariance.

The idea that dynamical state reduction might be induced from one field to another is not new. It has
appeared in previous attempts to describe state reduction in fermionic field theory by invoking a scalar 
field with state reduction properties to which the fermion is coupled \cite{pear3,ghir1}. To our knowledge 
the process has not been previously studied in numerical detail.

In the next section we outline our model, detailing the Hamiltonian for our simplified QED system and 
the modified stochastic dynamics of the state vector. In section \ref{sec3} we consider the evolution
of a specific initial state vector composed of a superposition of current states. We outline the state
reduction process and derive a characteristic time-scale. In section \ref{sec4} we present some numerical 
results which confirm the analytic results and intuition gained in section \ref{sec3}. We end with some conclusions.

\section{The model}

\subsection{QED Hamiltonian}

Our simplified QED Hamiltonian is given by
\bea
H = \omega a^{\dagger}a +\half\nu\sigma_z + g \sigma_x (a+a^{\dagger}).
\eea
The $\sigma_i$s are the standard Pauli matrices and $a$, $a^{\dagger}$ are the annihilation and creation 
operators for a standard harmonic oscillator. The parameters $\omega$, $\nu$, and $g$ are constants
which we interpret as the field mode frequency, the two-state energy gap, and the coupling respectively.

Models of this type are commonly used in quantum optics to describe the interaction of a two-state atom with 
an electromagnetic field \cite{optics2}. In particular, a further approximation to the interaction term to exclude processes such 
as those where a photon is created as the atom is exited (the rotating wave approximation), leads to the celebrated 
Jaynes-Cummings model. Alternatively, this model can be derived from full QED by restricting space to just two 
points (time remains continuous) and considering only one photon polarisation. The result is two fermion states 
and one field mode. This simplification is inadequate for a number of reasons. Most importantly, the model can no longer 
be described as a gauge theory as there is no local symmetry associated with the fermion field. We also lose the 
concepts of conserved 4-current and relativistic invariance. 

On the other hand, our simplified Hamiltonian has the structure of the QED Hamiltonian: it is composed of
free matter and field parts and an interaction between a current operator $\sigma_x$ and the field. 
(That $\sigma_x$ corresponds to the current can be seen as we restrict full QED to a space of two points.) 
These features suffice in order to make a tentative attempt to understand quantum state 
reduction in QED.

Since this model can be applied to quantum optical systems, it is worth pointing out that stochastic adjustments to
Schr\"odinger dynamics have been used in the past to describe the effect of an environment on optical and atomic 
behaviour \cite{carm,perci,knight}.

\subsection{Stochastic Schr\"odinger equation}

We shall consider a modified Schr\"odinger equation of the form
\bea
\rd|\psi_t\rangle = \left\{\left(- i H -\lambda^2 a^{\dagger}a 
+\lambda^2 a^{\dagger}_t a - \half\lambda^2 a^{\dagger}_t a_t	\right)\rd t
+\lambda\left(a-\half a_t \right) \rd B_t   -\half\lambda a^{\dagger}_t \rd B^{*}_t\right\}
|\psi_t\rangle,
\label{sse1}
\eea
where $\lambda$ is a constant coupling parameter with units of $[\rm time]^{-1/2}$ and 
\bea
a_{t} = \langle\psi_t|a|\psi_t\rangle.
\eea
The complex Wiener process $B_t$ satisfies 
$\rd B_t \rd B^*_t= 2\rd t $, and $\rd B_t \rd B_t = \rd B^*_t \rd B^*_t=0$ \cite{HIDA}.
Note that the Schr\"odinger equation is recovered by setting $\lambda=0$ and that the additional 
$\lambda$-dependent terms only involve field operators. 

Equation (\ref{sse1}) is a stochastic differential equation for the state.
It can be derived by looking for a state diffusion process describing stochastic 
fluctuations in the quantum probability amplitudes of possible $a$-eigenstates \cite{Bass}. The probabilities 
for the stochastic fluctuations must be given by the quantum probabilities at each point in time. 
This ensures that stochastic outcomes match with the initial quantum predictions (and leads to 
nonlinearity). The equation has the property that general states evolve towards $a$-eigenstates on a 
time-scale approximately inversely proportional to the variance in $a$ (see below). 

Using the rules of It\^o calculus we can show that 
\bea
\rd \langle\psi_t|\psi_t\rangle = 0,
\eea
i.e. the norm of the state is preserved under the dynamics (\ref{sse1}). This allows us to maintain the quantum interpretation of 
the state as a probability amplitude.

The process for the conditional expectation of $a$ is found to be
\bea
\rd a_t = \left\{-i\omega a_t - i g \langle \sigma_x\rangle_t - \lambda^2 a_t\right\}\rd t
+\lambda\langle (\Delta a)^2\rangle_t\rd B +\lambda\langle |\Delta a|^2\rangle_t\rd B^{\dagger},
\label{aproc}
\eea
where $\Delta a = a -a_t $ and we have used the convenient notation $\langle\cdot \rangle_t = \langle\psi_t|\cdot|\psi_t\rangle$.
When the state $|\psi_t\rangle$ is an $a$-eigenstate, the process $a_t$ is nonstochastic. 
In this case, the effect of the modifications to Schr\"odinger's equation is a decay in the magnitude of 
$a_t$ at a rate given by $\lambda^2$. This can be interpreted as being due to the fact that the classical 
stochastic process $B_t$ in equation (\ref{sse1}) is only coupled to photon annihilation events. This results 
in a gradual loss of energy. Appropriate tuning of $\lambda$ is required to ensure that field energy loss 
is negligible whilst state reduction timescales are appropriately small for strong fields \cite{me}. 

In order to demonstrate the state reducing properties of equation (\ref{sse1}) we consider the conditional 
variance in $a$ defined by
\bea
{\rm Var}_t(a) = \langle|\Delta a|^2\rangle_t
\eea
The conditional variance process is found to be
\bea
\rd {\rm Var}_t(a) &=& i g \left\{ \langle\sigma_x(a-a^{\dagger})\rangle_t 
				- \langle \sigma_x \rangle_t \langle (a-a^{\dagger}) \rangle_t \right\}\rd t
\nonumber\\
&& -2\lambda^2{\rm Var}_t(a) \rd t -2\lambda^2\left\{ |\langle (\Delta a)^2 \rangle_t|^2 + {\rm Var}_t^2(a)\right\}\rd t
\nonumber\\
&& +\lambda \langle |\Delta a|^2\Delta a \rangle_t \rd B_t 
+ \lambda \langle \Delta a^{\dagger} |\Delta a|^2\rangle_t\rd B^*_t .
\label{var}
\eea
The first line on the right side is the effect on ${\rm Var}_t(a)$ due to the interaction between the two-state 
system and field. We see that ${\rm Var}_t(a)$ increases if there is a negative conditional covariance 
between $\sigma_x$ and $ -i(a-a^{\dagger}) $. As ${\rm Var}_t(a)$ increases, other terms on the right side of
equation (\ref{var}) become more important. 

If we ignore the interaction with the two-state system the variance
process is a super-martingale. Integrating and taking the unconditional expectation of equation (\ref{var}) 
we then find
\bea
\mathbb{E}[{\rm Var}_t(a)] ={\rm Var}_0(a) -2\lambda^2\int_0^t\rd u \mathbb{E}[{\rm Var}_u(a)]
-2\lambda^2\int_0^t\rd u \mathbb{E}\left[|\langle (\Delta a)^2 \rangle_u|^2 + {\rm Var}_u^2(a)\right].
\label{varexp}
\eea
This demonstrates that the expected variance will decrease in time and therefore that the field approaches 
an $a$-eigenstate. 

We can estimate the time-scale $\tau_a$ for state reduction to an $a$-eigenstate from the following:
\bea
\frac{\mathbb{E}[{\rm Var}_t(a)] - {\rm Var}_0(a)}{{\rm Var}_0(a)} \sim \frac{t}{\tau_a}.
\eea   
By freezing the stochastic terms on the right side of equation~(\ref{varexp}) at time $t=0$ we find
\bea
\tau_a \sim \frac{{\rm Var}_0(a)}{\lambda^2 \left\{|\langle (\Delta a)^2 \rangle_0|^2 + {\rm Var}_0^2(a)\right\}}.
\label{timescale}
\eea 
Even though $\lambda$ must be very small to prevent energy loss, the reduction time-scale $\tau_a$ can 
be small if the variance in $a$ is large. This demonstrates the specific way in which the effectiveness of 
state reduction is dependent on the size of the system. A superposition of greatly differing $a$-eigenstates
has large variance and state reduction will occur rapidly; if ${\rm Var}_0(a) \sim |\langle (\Delta a)^2 \rangle_0| \sim {\cal O}(1)$ 
then state reduction takes an extremely long time due to the small value of $\lambda$.
In principle, the dependence of the reduction time on the $a$-variance is a testable result. An experiment 
designed to test this might look for diminishing quantum interference as a signature of the state reduction \cite{legg}. 
The technical difficulty would be to eliminate any environmental influence.

In order to see that stochastic probabilities for outcomes match with the initial quantum estimates
we consider an initial superposition of coherent states
\bea
|\psi_0\rangle = \sum_i c_i |\alpha_i\rangle,
\eea
where it is supposed that the coherent states $|\alpha_i\rangle$, where 
$a|\alpha_i\rangle=\alpha_i|\alpha_i\rangle$, are sufficiently separated in phase space that 
$\langle\alpha_i|\alpha_j\rangle\simeq\delta_{ij}$ (this requires that $|\alpha_i-\alpha_j|\gg 1$ for $i\neq j$).
We assume for simplicity that $H=0$ since we wish to focus on the reduction dynamics.

Note that the coherent states are eigenstates of all terms in the evolution equation~(\ref{sse1}) except
the term $-\lambda^2a^{\dagger}a$. This means that most terms in the evolution equation simply modify the 
coefficients $c_i$, leaving the set of states $\{|\alpha_i\rangle\}$ intact. The term $-\lambda^2a^{\dagger}a$ 
is responsible for diffusing the initial set of coherent states.
In order to analyse its effects, consider a state evolution involving just this term:
\bea
\frac{\rd}{\rd t}|\psi_t\rangle = -\lambda^2 a^{\dagger}a |\psi_t\rangle .
\eea
This equation has the solution
\bea
|\psi_t\rangle = \sum_i c_i \exp \left\{\half(|\alpha_i e^{-\lambda^2 t}|^2-|\alpha_i|^2)\right\}
|\alpha_i e^{-\lambda^2 t}\rangle.
\eea
This indicates that the individual coherent state contributions to the overall superposition 
will decay to the ground state on a timescale $\lambda^{-2}$. For times $t\ll \lambda^{-2}$ 
we can ignore this effect and assume that the set of coherent states contributing to the 
superposition is fixed as $\{|\alpha_i\rangle\}$. From equation~(\ref{timescale}) we see that 
$\tau_a\ll\lambda^{-2}$ for large $a$-variance. We can therefore assume that in the region 
$\tau_a \ll t \ll \lambda^{-2}$, the state has reduced to one of the coherent states $|\alpha_i\rangle$.

We can define an approximate coherent state projection operator by $P_{\alpha_k}=|\alpha_k\rangle\langle\alpha_k|$. 
For $\tau_a \ll t \ll \lambda^{-2}$ the conditional expectation
$\langle\psi_t|P_{\alpha_k}|\psi_t\rangle$ is approximately either 1 or 0 depending on whether the state is reduced
to $|\alpha_k\rangle$ or not. From equation (\ref{sse1}) we find
\bea
\rd\{\langle\psi_t|P_{\alpha_k}|\psi_t\rangle\}
=\lambda|c_k(t)|^2(\alpha_k-a_t)\rd B_t
+\lambda|c_k(t)|^2(\alpha_k^*-a_t^{\dagger})\rd B^*_t,
\eea
from which we conclude that 
\bea
\langle\psi_0|P_{\alpha_k}|\psi_0\rangle = \mathbb{E}[\langle\psi_t|P_{\alpha_k}|\psi_t\rangle|{\cal F}_0]
=\mathbb{E}[\mathbf{1}_{|\psi_t\rangle=|\alpha_k\rangle}|{\cal F}_0],
\eea
valid for $\tau_a \ll t \ll \lambda^{-2}$. The left hand equality gives the initial quantum probability for the 
outcome $|\psi_t\rangle=|\alpha_k\rangle$; the right hand equality gives the stochastic probability for the same outcome.

\section{An experiment}
\label{sec3}

Consider an initial situation where the field is in a coherent state $|\alpha\rangle$ and the two-state system is 
in a superposition of current ($\sigma_x$) eigenstates: 
\bea
|\psi(0)\rangle = \left\{ c_1 \cura + c_2\curb \right\} |\alpha\rangle.
\label{init}
\eea
In order to understand the behaviour of this model without explicitly solving it we consider some possible scenarios 
depending on the relative sizes of the model parameters. 

Since the field state is initially coherent, the variance in $a$ is initially small. As this is our measure of system size, 
we can assume that equation (\ref{sse1}) initially reduces to the Schr\"odinger equation.

\subsection{$\omega$, $\nu$ domination}
First consider the case where the coupling $g$ tends to zero. We then have
\bea
|\psi(t)\rangle_{g=0} &=& \exp\left\{ -i t \left(\omega a^{\dagger}a +\half\nu\sigma_z\right)\right\}|\psi(0)\rangle
\nonumber\\
&=&\left\{c_1(t) \cura
		+c_2(t) \curb\right\}
		|{\rm e}^{-it\omega}\alpha\rangle,
\eea 
where
\bea
c_1(t) &=& c_1\cos\half t\nu-ic_2\sin\half t\nu,\nonumber\\
c_2(t) &=& c_2\cos\half t\nu-ic_1\sin\half t\nu.
\eea
Here we find independent state-space rotations for the two-state system and the field.

\subsection{$g$ domination}
Next consider the case where the field frequency $\omega$ and the two-state energy gap $\nu$ tend to zero. In this 
limit we find the following solution to the Schr\"odinger equation  
\bea
|\psi(t)\rangle_{\omega = \nu = 0} &=& \exp\left\{ -i t g \sigma_x (a+a^{\dagger})\right\}|\psi(0)\rangle
\nonumber\\
&=& c_1 e^{i\phi_1} \cura |(\alpha-igt)\rangle + c_2 e^{i\phi_2} \curb|(\alpha+igt)\rangle,
\label{gdom}
\eea
where $\phi_i$ are real-valued phase factors.
The interaction between two-state system and field leads to a correlation between current eigenstates and field 
coherent states.

These two limits provide an overall picture of two competing processes. When the interaction $g$ is dominant, 
different current eigenstates become associated with different coherent field states. When the interaction is negligible
we find that current states rotate into one another.

\subsection{$\lambda$ domination}
Consider the scenario $\omega\sim\nu\ll g$. From equation (\ref{gdom}) we can estimate the growth of variance in $a$:
\bea
{\rm Var}_t(a)  \simeq 4 g^2 t^2 |c_1|^2|c_2|^2 \simeq |\langle (\Delta a)^2 \rangle_t| ,
\label{vargrow}
\eea
and the conditional covariance between $\sigma_x$ and $-i(a-a^{\dagger})$:
\bea
{\rm Cov}_t\left(\sigma_x, -i(a-a^{\dagger})\right) &=& -i\langle\sigma_x(a-a^{\dagger})\rangle_t 
				+i \langle \sigma_x \rangle_t \langle (a-a^{\dagger}) \rangle_t 
\nonumber\\
&\simeq& -8 g t |c_1|^2|c_2|^2 
\label{covgrow}
\eea
(the approximation results from the fact that we have assumed that different coherent states are nearly orthogonal).
As the variance grows, $\lambda$-effects come to dominate the evolution. This can be seen from equations
(\ref{vargrow}) and (\ref{covgrow}) by noting that the ${\rm Var}^2_t(a)$ and $ |\langle (\Delta a)^2 \rangle_t|^2$ 
terms in equation (\ref{var}) will soon outgrow the ${\rm Cov}_t\left(\sigma_x, -i(a-a^{\dagger})\right)$ term.
Let us define an approximate time $s$ at which a transition from $g$ to $\lambda$-dominated behaviour happens. 
Once in the $\lambda$-dominated phase, the $a$-variance will tend to zero 
on a time-scale given by equation (\ref{timescale}). The stochastic evolution effectively chooses between the possible states 
$(1,\;1)^{\dagger}|\alpha-igt\rangle$ and $(1,\;-1)^{\dagger}|\alpha+igt\rangle$ leading towards a definite field 
coherent state and a definite current state.

Using (\ref{timescale}) and (\ref{vargrow}) we find that a total state reduction time-scale for the current state
is given by
\bea
\tau_{\sigma_x} \sim s + \frac{1}{\lambda^2 g^2 s^2 |c_1|^2|c_2|^2 }.
\eea 
In order to determine $s$ we find the value which minimises $\tau_{\sigma_x}$. Setting
\bea
\frac{\rd}{\rd s} \tau_{\sigma_x} = 0 \implies 1 \sim \frac{1}{\lambda^2g^2 s^3|c_1|^2|c_2|^2 } , 
\eea
therefore
\bea
\tau_{\sigma_x} \sim \frac{1}{\lambda^{2/3}g^{2/3}|c_1|^{2/3}|c_2|^{2/3}}.
\label{indred}
\eea
This is the characteristic time-scale for state reduction of a superposition of current eigenstates in a 
two-state system coupled to a field mode.

\section{Numerical results}
\label{sec4}
\begin{figure}[t]
\label{figvar}
\begin{center}
\includegraphics[width=10cm]{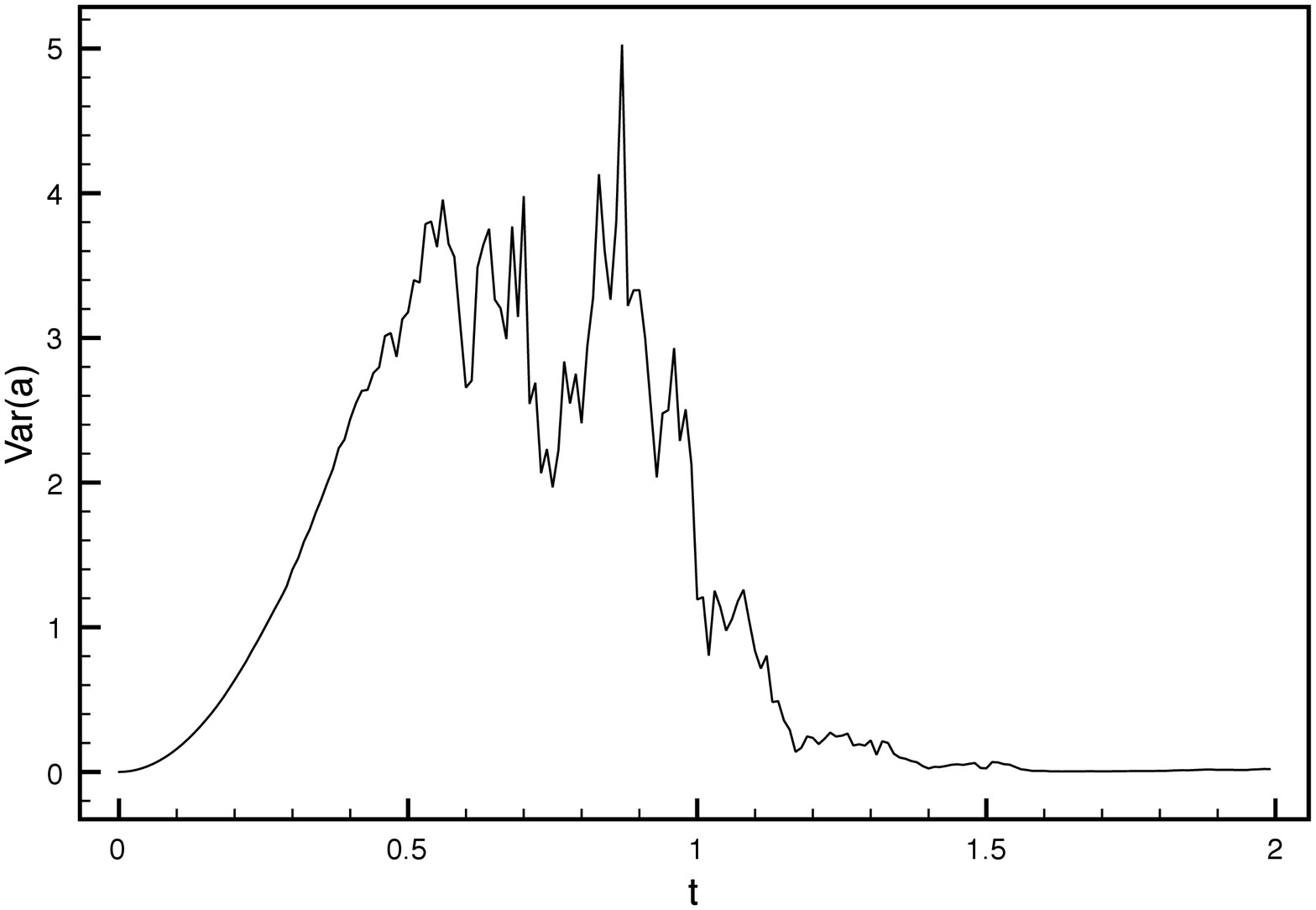}
\caption{\textsf{Sample path showing growth and subsequent decay of ${\rm Var}_t(a)$.
The inital state involves an equal superposition of two different current states. The coupling
between current and field mode leads at first to a growth in ${\rm Var}_t(a)$ via standard Hamiltonian 
dynamics. When ${\rm Var}_t(a)$ is sufficiently large, the stochastic dynamics cause a decay in variance 
as the $a$-state collapses. Parameters are $\omega=\nu=0.5$, $g=4.0$, $\lambda=0.2$.}}
\end{center}
\end{figure}

\begin{figure}[t]
\label{figsig}
\begin{center}
\includegraphics[width=10cm]{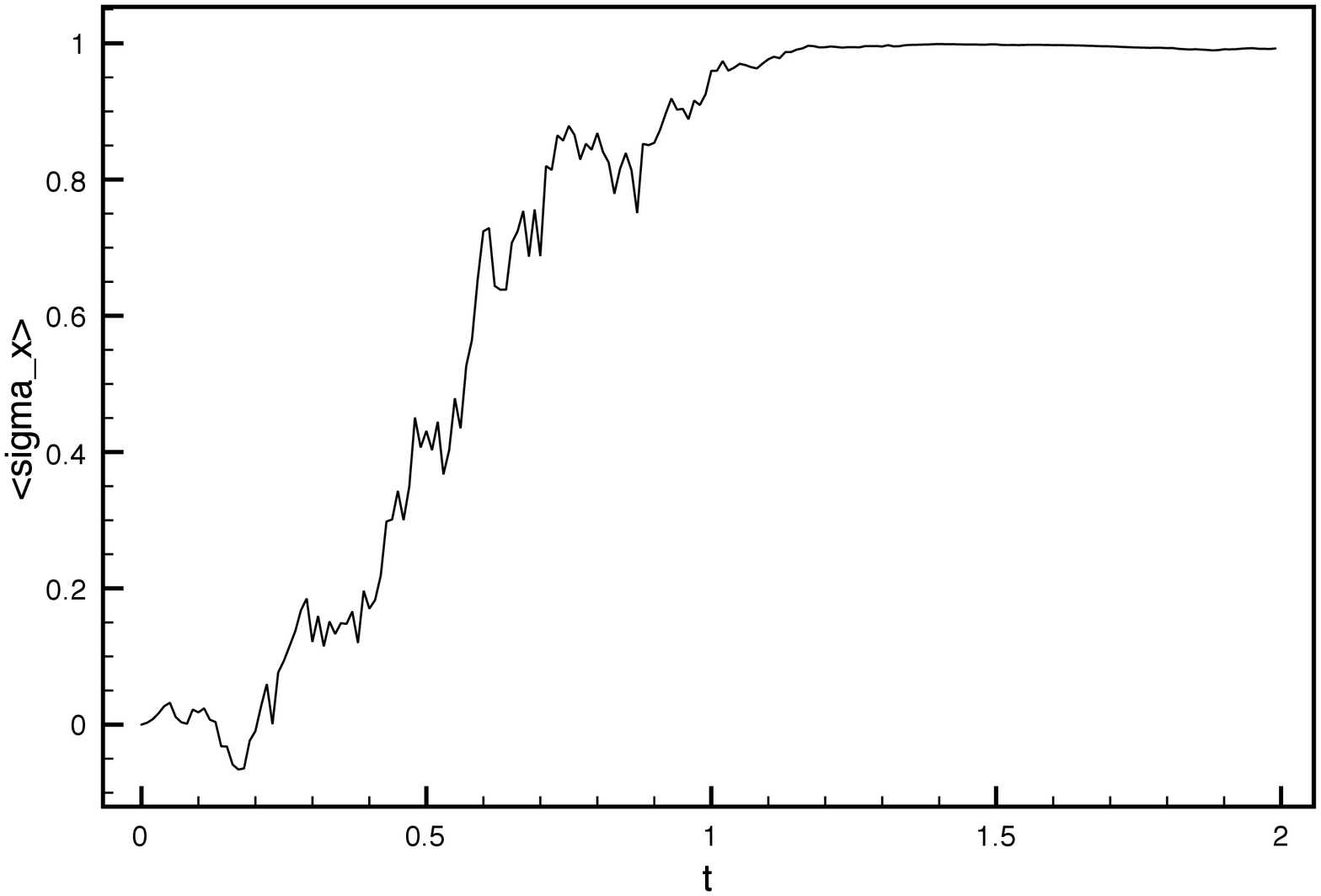}
\caption{\textsf{Evolution of the expected current $\langle\sigma_x\rangle_t$. An initial (equal) superposition
of two different current states ($\langle\sigma_x\rangle_0=0$) undergoes state reduction to a definite current 
state ($|\langle\sigma_x\rangle_t|=1$). Same random sample as figure 1.}}
\end{center}
\end{figure}

In order to test these results we have implemented a numerical simulator. Starting from an initial state
of the form given in equation (\ref{init}) with coefficients $c_1=c_2=1/2$ and with $\alpha=4$, we have
evolved the state according to equation (\ref{sse1}). Figures 1 and 2 show sample paths for ${\rm Var}_t(a)$
and $\langle\sigma_x\rangle_t$ respectively. Here we have chosen $\omega=\nu=0.5$, $g=4.0$, $\lambda=0.2$.
This choice of parameters are such that $g$ dominates the initial evolution. The parameter $\lambda$ is chosen to be large enough
to observe state reduction for the modestly sized numerical system, but small enough that energy loss is not too 
significant.

It is seen that the $a$-variance initially grows smoothly, approximately proportional to $t^2$ (see equation 
(\ref{vargrow})). After a certain time the evolution becomes dominated by stochastic movements with a notable 
downward drift. The $a$-variance finally becomes stable again as its value tends to zero (at around $t=1.5$). 

The current in figure 2 corresponds to the same sample. It is seen that the two-state system tends stochastically towards 
a definite current state---in this particular case $(1,\;1)^{\dagger}$. This happens at the same time as the $a$-variance 
goes to zero.

\begin{figure}[ht]
\label{figstop}
\begin{center}
\includegraphics[width=10cm]{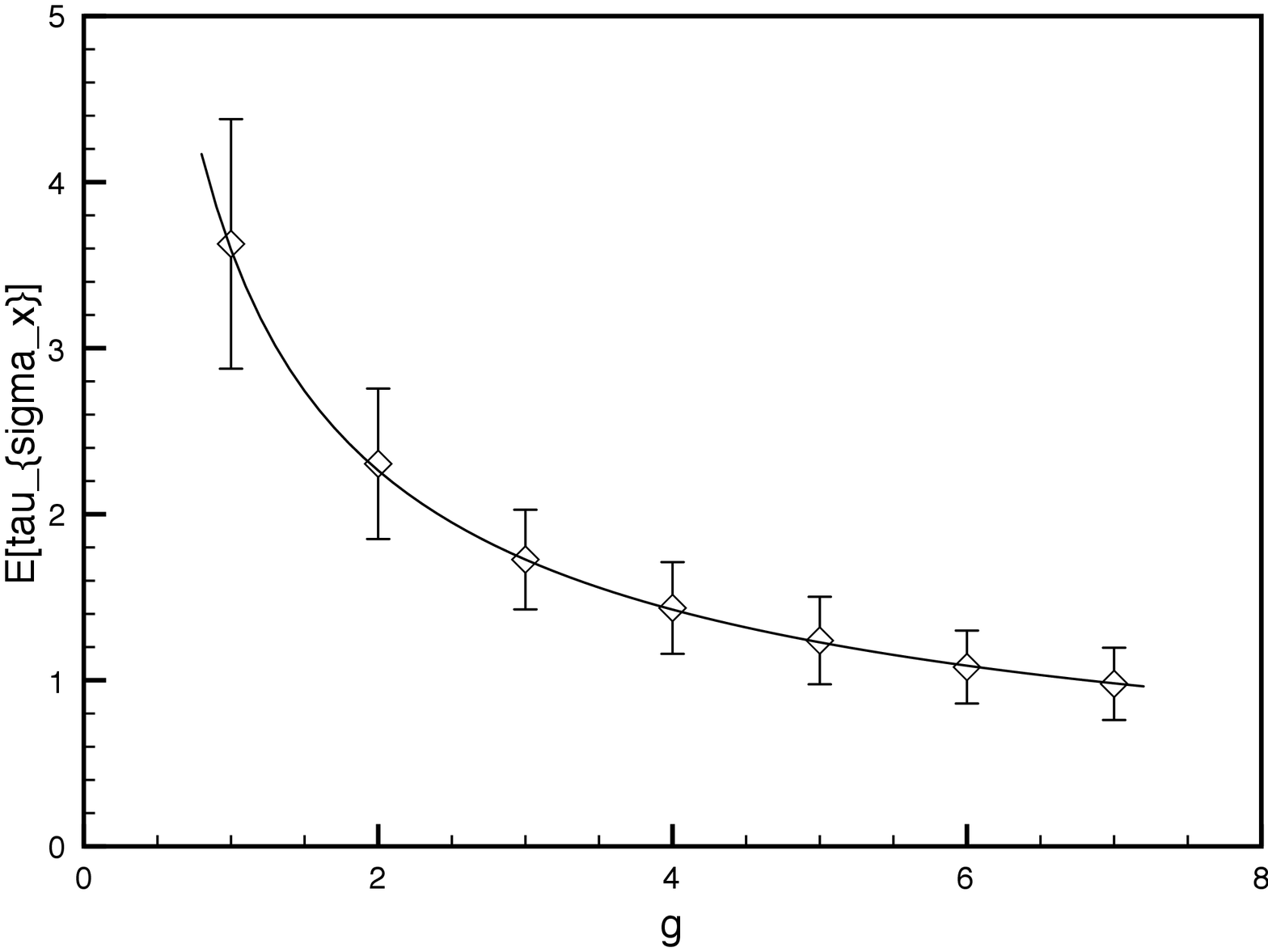}
\caption{\textsf{Expected current reduction time as a function of the current-field coupling $g$. 
Other parameters are $\omega=\nu=0$, $\lambda=0.2$. The filled line shows a fit to the 
curve $g^{-2/3}$. The error bars show the one standard deviation region for the reduction time 
estimated from the simulated data.}}
\end{center}
\end{figure}

\begin{figure}[ht]
\begin{center}
\includegraphics[angle=-90,width=10cm]{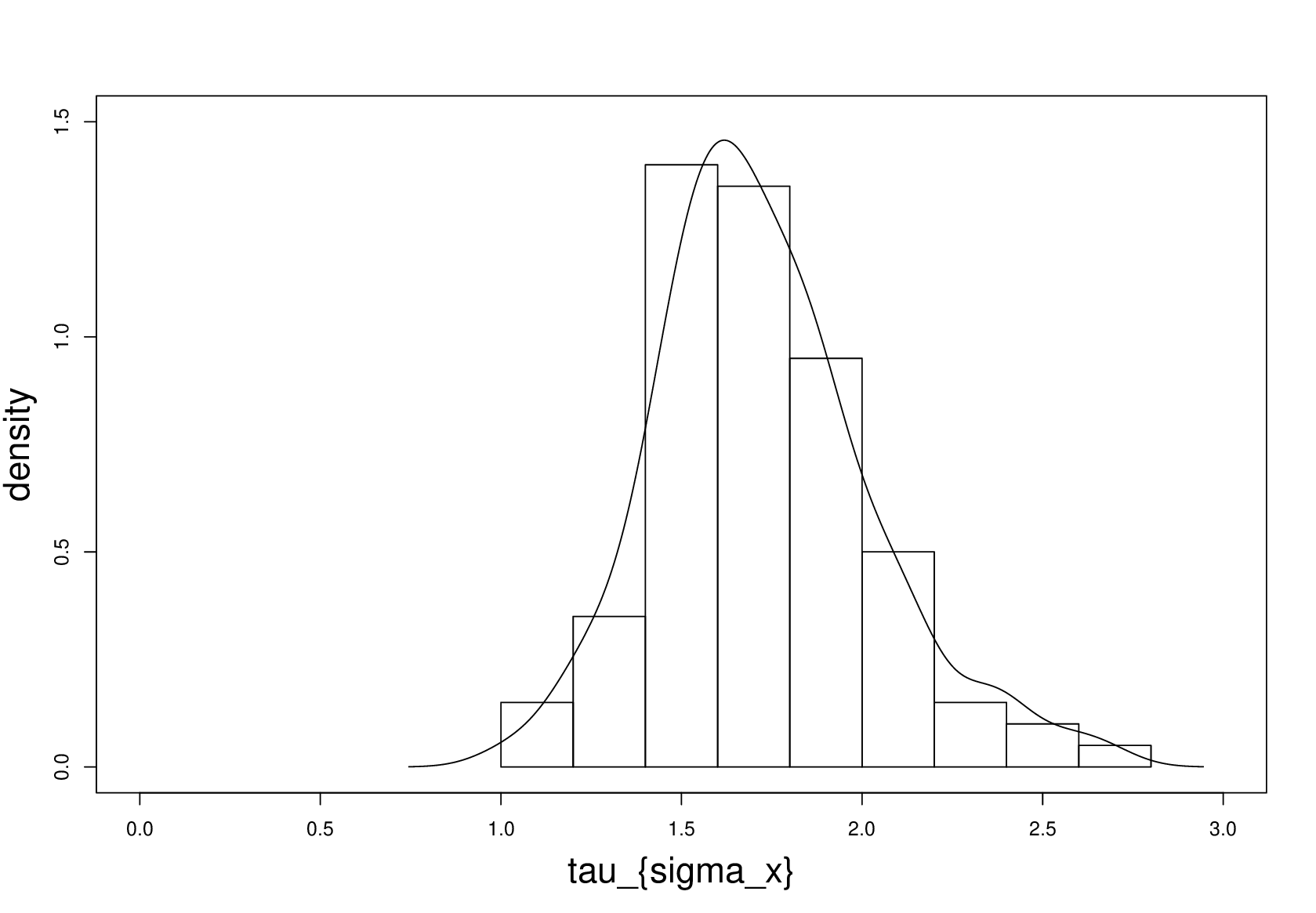}
\caption{\textsf{Histogram of $|\langle\sigma_x\rangle_t|>0.99$ stopping times for $g=3$ 
($\omega=\nu=0$, $\lambda=0.2$). The smoothed line is the Gaussian kernel density estimate.}}
\end{center}
\end{figure}

We define the reduction time in our numerics as the stopping time at which $|\langle\sigma_x\rangle_t|>0.99$, i.e.
\bea
\tau_{\sigma_x} = {\rm inf}\left\{t>0;|\langle\sigma_x\rangle_t|>0.99 \right\}.
\eea
Figure 3 shows numerical estimates of the expectation of this stopping time for a range of $g$ values. Other parameters are chosen to be
$\lambda=0.2$ and $\omega=\nu=0$. Expectations are determined using 100 sample paths for each point. The estimates
are seen to almost perfectly fit the curve $kg^{-2/3}$ with $k=3.593$. This confirms the form of the characteristic state reduction time 
given in equation (\ref{indred}). The error bars indicate a range of one standard deviation for $\tau_{\sigma_x}$. We observe 
that the variance in the stopping time decreases as $g$ increases. 

Figure 4 shows a histogram of $\tau_{\sigma_x}$ results for 100 sample paths with $g=3$. 
Also shown is the Gaussian kernel density estimate \cite{kernel}. The distribution of stopping times is approximately normal 
with a slight skew. The important result is that the density of events with large deviation from the mean stopping time 
falls away rapidly. 

We find that occurrences of $\langle\sigma_x\rangle_t\rightarrow 1$ and $\langle\sigma_x\rangle_t\rightarrow -1$ 
are even to within statistical error. This agrees with standard quantum predictions for a measurement of $\sigma_x$ 
given the initial conditions.

\section{Conclusions}
We have investigated the state reduction dynamics described by a modified Schr\"odinger equation for a simplified 
QED system.  Our simplified QED system is composed of a charged two-state system interacting 
with an electromagnetic field mode. The modified dynamics result in state reduction to an over-complete coherent 
field state basis. At the same time, the electromagnetic interaction leads to the development of correlation between 
current states and field coherent states. This leads to an induced state reduction to a current state basis for the 
two-state system. We have considered evolution from a specific initial condition composed of an equal superpostion 
of the two possible current eigenstates with a coherent field state. We have implemented a numerical simulator to 
generate sample state evolutions. These have confirmed our analytic predictions for the general behaviour and the 
time-scales involved.

The time-scale for state reduction of the the two-state system is found to behave as $g^{-2/3}{\rm Var}^{-1/3}(\sigma_x)$. 
We might speculatively suggest that the reduction time-scale for more general charged matter systems might scale 
as ${\rm Var}^{-1/3}(j_{\mu})$ for 4-current $j_{\mu}$.

\section*{Acknowledgements}
I would like to thank Philip Pearle and Dorje Brody for useful comments.

\end{document}